\documentclass[reprint,superscriptaddress,aps,prr]{revtex4-2}
\usepackage{amsmath}
\usepackage{graphicx}
\usepackage{subfigure}
\usepackage{amssymb}

\begin{document}
	\title{Improving D2p Grover's algorithm to reach performance upper bound under phase noise}
	
	\author{Jian Leng}
	\affiliation{ State Key Laboratory of Low Dimensional Quantum Physics, Department of Physics, \\ Tsinghua University, Beijing 100084, China}
	\author{Fan Yang}
	\affiliation{ State Key Laboratory of Low Dimensional Quantum Physics, Department of Physics, \\ Tsinghua University, Beijing 100084, China}
	\author{Xiang-Bin Wang}
	\email{ xbwang@mail.tsinghua.edu.cn}
	\affiliation{ State Key Laboratory of Low Dimensional Quantum Physics, Department of Physics, \\ Tsinghua University, Beijing 100084, China}
	\affiliation{ Jinan Institute of Quantum technology, SAICT, Jinan 250101, China}
	\affiliation{ Shanghai Branch, CAS Center for Excellence and Synergetic Innovation Center in Quantum Information and Quantum Physics, University of Science and Technology of China, Shanghai 201315, China}
	\affiliation{ Shenzhen Institute for Quantum Science and Engineering, and Physics Department, Southern University of Science and Technology, Shenzhen 518055, China}
	\affiliation{ Frontier Science Center for Quantum Information, Beijing, China}
	
	\begin{abstract}
		The original Grover's algorithm has a success probability to output a correct solution, while deterministic Grover's algorithms improve the success probability to $100\%$. However, the success probability of deterministic Grover's algorithm decreases in noisy environment. Here we improve the deterministic two-parameter (D2p) Grover's algorithm to reach the upper bound for success probability under phase noise. We prove that it is not possible to design any deterministic Grover's algorithm whose success probability is higher than our improved D2p protocol's under phase noise.
	\end{abstract}
	
	
	\maketitle
	
	\section{Introduction}
	
	Grover's quantum search algorithm \cite{grover1996fast,grover1997quantum} can effectively solve the search problem which has total $N$ inputs and $M$ solutions. It gives a correct solution with $\mathcal{O}(\sqrt{N/M})$ operations, and its success probability is better than $50\%$. This provides a quadratic speedup against classical search algorithms. Grover's algorithm can be described as the iterations of two modules: the black box oracle which is untunable \cite{Nielsen2002} and the reflection operator with a tunable phase $\beta$. The original Grover's algorithm simply chooses $\beta=\pi$. Very recently, a novel result named as deterministic two-parameter (D2p) Grover's algorithm \cite{roy2022deterministic} is presented. This D2p protocol can deterministically give a correct solution by choosing two well-designed phases $\beta_1,~\beta_2$ for reflection operator. There is an important problem that noise \cite{galindo2000family,shapira2003effect,salas2007robustness,salas2008noise,reitzner2019grover} is inevitable for practical quantum circuit. It is meaningful to find such a robust algorithm with success probability as high as possible. In particular, phase noise $\delta\beta$ is caused by implement's imperfection and, to our knowledge, designing a quantum error correction code for phase noise is still an open problem. It should be clarified that the phase noise $\delta\beta$ (coherent) appears in the logical tunable phase $\beta$ and it is different from the dephasing noise (incoherent) for which a few quantum error correction code exists. The effect of phase noise \cite{shenvi2003effects} to the original Grover's algorithm has been investigated. However, the optimal result of Grover's algorithm is unknown. Here we improve the D2p algorithm to reach the upper bound for success probability under phase noise. We prove that the success probability of our improved D2p protocol is the highest among all possible deterministic Grover's algorithms, which cannot be surpassed in principle.
	
	We shall use geometric method to study phase noise effect. Throughout this work, we consider the standard black box oracle without any user-tunable parameters \cite{Nielsen2002,roy2022deterministic} and do not include the type of oracle with user-tunable parameters \cite{long2001grover,toyama2013quantum}.
	
	The success probability criteria has important application~\cite{roy2022deterministic,long2001grover}. Especially in the cases such as high cost of system initialization and quantum measurement. Another important criteria is runtime~\cite{boyer1998tight,hamann2022performance} which maybe used in other literature.

	\section{The original and D2p Grover's algorithms}
	
	We begin with initial state $|\psi_0\rangle=\sum_{x=0}^{N-1}|x\rangle$, where the dimension of Hilbert space is $N=2^n$ and $n$ is the number of qubits. There are $M$ target states which are denoted as $|t_j\rangle$, and the others are written as $|r_j\rangle$. Then the initial state can be expressed in a two dimension space
	\begin{equation}
		|\psi_0\rangle=\sqrt{1-\lambda}|R\rangle+\sqrt{\lambda}|T\rangle:=
		\left(
			\begin{array}{c}
				\sqrt{1-\lambda}\\
				\sqrt{\lambda}
			\end{array}
		\right),
	\end{equation}
	where
	\begin{eqnarray}
		|T\rangle=&&\frac{1}{\sqrt{M}}\sum_{j=1}^{M}|t_j\rangle,\nonumber\\
		|R\rangle=&&\frac{1}{\sqrt{N-M}}\sum_{j=1}^{N-M}|r_j\rangle,
	\end{eqnarray}
	and $\lambda=M/N$. The black box oracle and reflection operators can also be written in this two dimension space \cite{yoder2014fixed}:
	\begin{align}\label{S_matrix}
		S_o=&	\left(
					\begin{array}{cc}
						1 & 0\\
						0 & -1
					\end{array}
				\right),\nonumber\\
		S_r(\beta)=&\left(
						\begin{array}{cc}
							1-(1-e^{i\beta})\lambda &  (1-e^{i\beta})\sqrt{\lambda(1-\lambda)} \\
							(1-e^{i\beta})\sqrt{\lambda(1-\lambda)} & e^{i\beta}+(1-e^{i\beta})\lambda
						\end{array}
					\right),
	\end{align}
	and their product is the Grover's iterate operator:
	\begin{align}\label{G_operator}
		G(\beta)=&-S_r(\beta)S_o\nonumber\\
				=&	\left(
						\begin{array}{cc}
							(1-e^{i\beta})\lambda-1 &  (1-e^{i\beta})\sqrt{\lambda(1-\lambda)} \\
							-(1-e^{i\beta})\sqrt{\lambda(1-\lambda)} & e^{i\beta}+(1-e^{i\beta})\lambda
						\end{array}
					\right).
	\end{align}
	
\begin{figure}[tbp]
		\begin{minipage}{1\linewidth}
			\centering
			\includegraphics[width=0.8\linewidth]{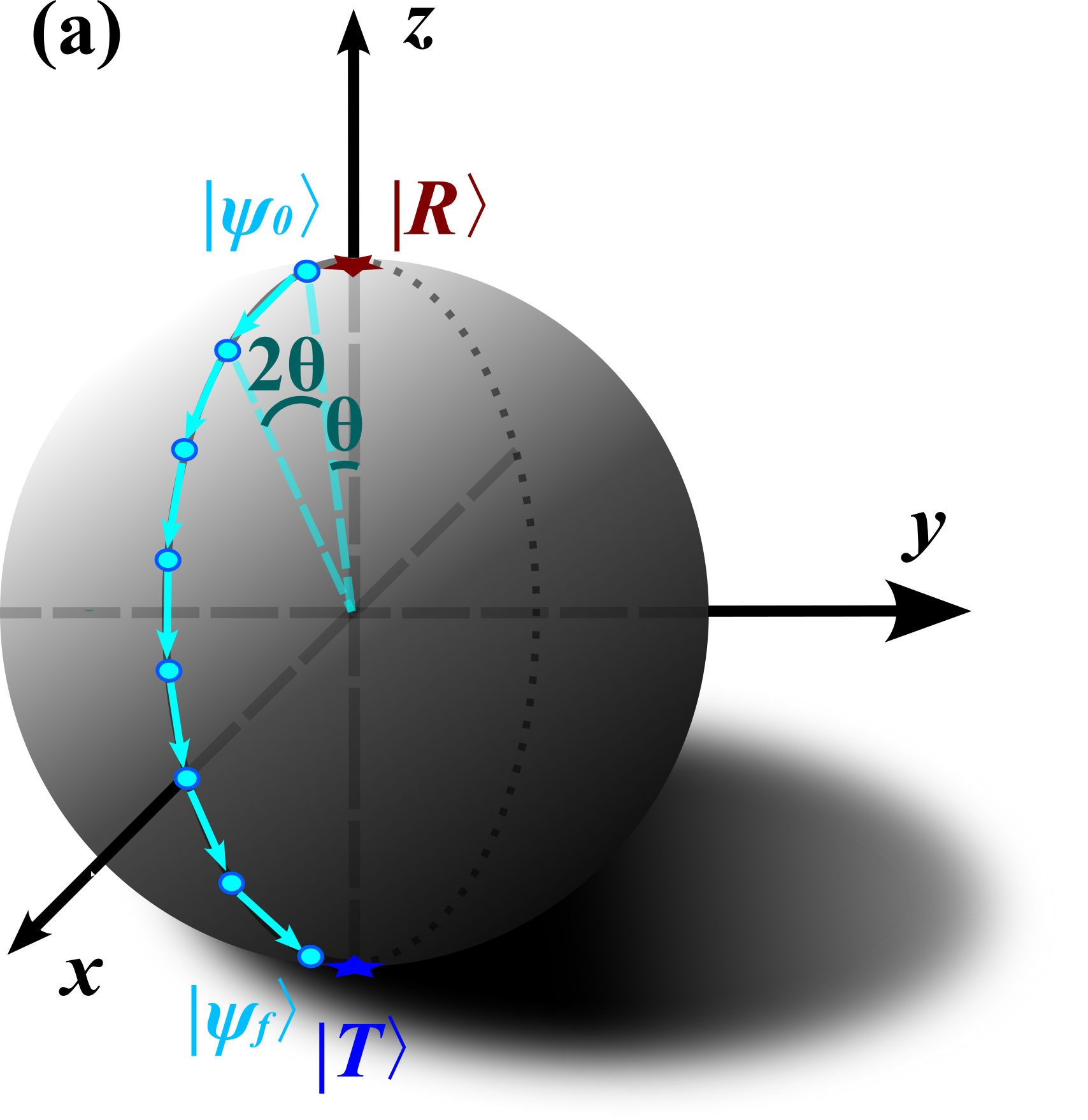}
		\end{minipage}
		
		\begin{minipage}{1\linewidth}
			\centering
			\includegraphics[width=0.8\linewidth]{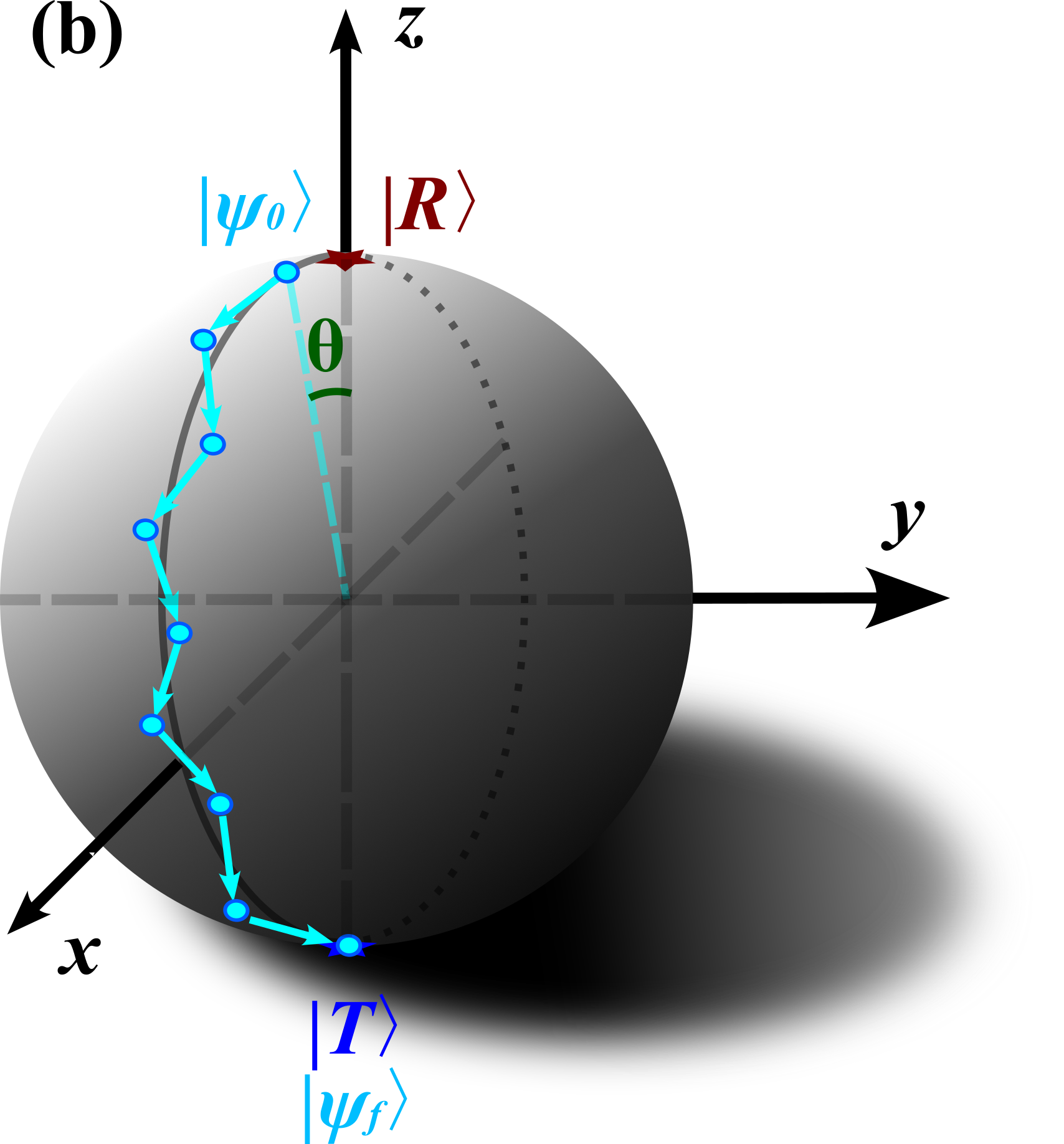}
		\end{minipage}
	
		\begin{minipage}{1\linewidth}
			\centering
			\includegraphics[width=1\linewidth]{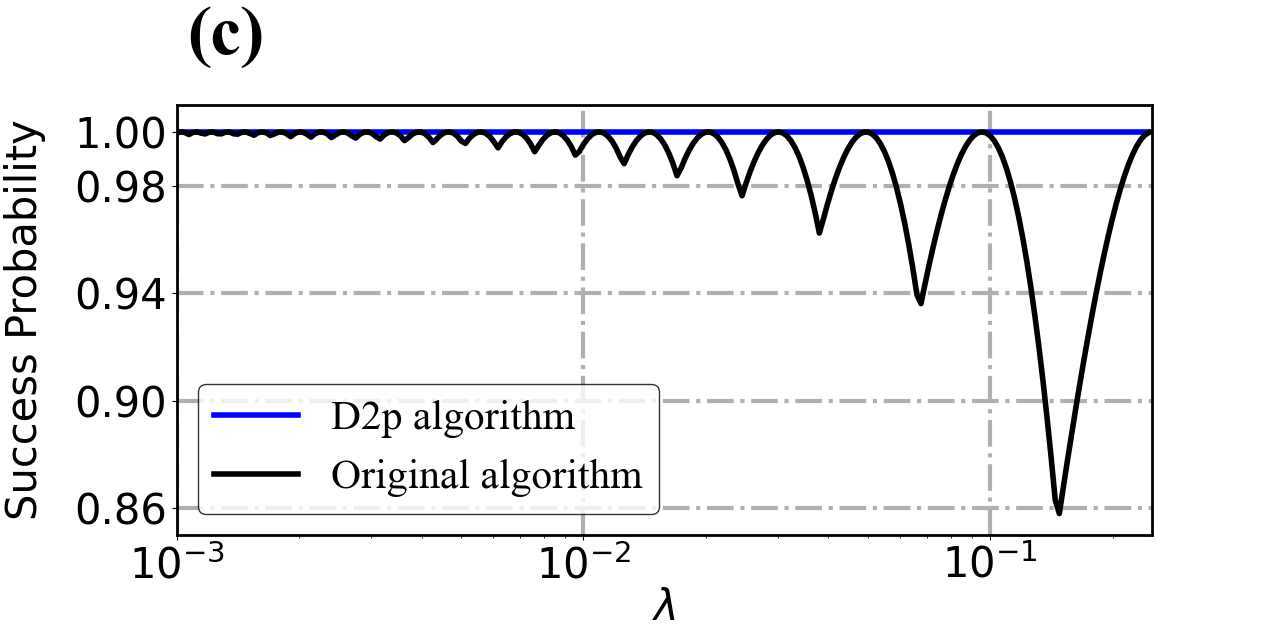}
		\end{minipage}
		
		\caption{\label{grover_d2p_review}The original Grover's algorithm and deterministic two-parameter (D2p) protocol. (a) On Bloch sphere, the original Grover's algorithm rotates initial state $|\psi_0\rangle$ about y axis step by step. Each step takes $2\theta$ angle. At the end of path, the final state $|\psi_f\rangle$ usually cannot coincide with target state $|T\rangle$. (b) D2p protocol has two different Grover's iterate operators $G(\beta_1)$ and $G(\beta_2)$. These two operators alternatively rotate initial state $|\psi_0\rangle$ on Bloch sphere and draw a zigzag line. The final state $|\psi_f\rangle$ exactly coincides with target state $|T\rangle$. (c) These two algorithms are simulated in noiseless environment, and success probability $|\langle\psi_f|T\rangle|^2$ are plotted. Clearly, D2p has better performance than original Grover's algorithm.}
	\end{figure}

	Both original and D2p algorithms repeatedly apply Grover's iterate operator $G(\beta)$ to initial state $|\psi_0\rangle$ with different $\beta$. The original algorithm chooses $\beta=\pi$, then $G(\pi)$ rotates $|\psi_0\rangle$ towards $|T\rangle$ along the Bloch sphere's geodesic step by step as shown in Fig. \ref{grover_d2p_review}(a). Each step rotates $2\theta=4\sin^{-1}\sqrt{\lambda}$, and after $k_0$ steps the state will exactly coincide with $|T\rangle$, where
	\begin{equation}
		k_0=\frac{\pi}{2\theta}-\frac{1}{2}=\frac{\pi}{4\sin^{-1}\sqrt{\lambda}}-\frac{1}{2}.\label{k_0}
	\end{equation}
	Usually $k_0$ is not an integer. We apply $k_g=\lfloor k_0\rceil$ steps to obtain a final state $|\psi_f\rangle$ which is closest to $|T\rangle$. So the original Grover's algorithm probably outputs a correct answer with success probability $|\langle\psi_f|T\rangle|^2=\sin^2[(k_g+1/2)\theta]$.
	
	D2p algorithm designs two phases $\beta_1, \beta_2$ for Grover's iterate operator, then $G(\beta_1)$ and $G(\beta_2)$ rotates $|\psi_0\rangle$ along a zigzag line on Bloch sphere as shown in Fig. \ref{grover_d2p_review}(b). In Ref. \cite{roy2022deterministic}, they show that
	\begin{equation*}
		\forall k_d\ge\lceil k_0\rceil,~\exists \beta_1,\beta_2,~\langle\psi_f|R\rangle=0.
	\end{equation*}
	It means that after $k_d$ steps the final state $|\psi_f\rangle$ deterministically coincides with $|T\rangle$. The phases $\beta_1$ and $\beta_2$ are given by solving the following equations with parameter $k_d$ and $\lambda$ \cite{roy2022deterministic}:
	\begin{eqnarray}
		1+4\lambda(1-2\lambda)\sin(\frac{\beta_1}{2})\sin(\frac{\beta_2}{2})\frac{\tan(\frac{k_d}{2}\phi)}{\sin\phi}=&&0,\nonumber\\
		(1-4\lambda)\tan(\frac{\beta_1}{2})+\tan(\frac{\beta_2}{2})=&&0,\nonumber\\
		\cos(\frac{\beta_1+\beta_2}{2})+8\lambda(1-\lambda)\sin(\frac{\beta_1}{2})\sin(\frac{\beta_2}{2})-\cos\phi=&&0.\nonumber\\
	\end{eqnarray}
	These equations work for even integer $k_d$. There are another set of equations when $k_d$ is odd \cite{roy2022deterministic}. The simulation results \cite{roy2022deterministic} for original and D2p algorithms are shown in Fig. \ref{grover_d2p_review}(c). It is apparent that one successes with probability smaller than $1$ and the other with certainty in a perfectly noiseless environment.
	
	\section{Phase noise}
	
	The original Grover algorithm chooses only one phase value $\beta=\pi$, and D2p protocol designs two different $\beta$ value \cite{Nielsen2002,roy2022deterministic}. In general Grover's algorithm, each step can take a different phase value $\beta$. Here we use geometrical property of Bloch sphere to prove that phase noise has different influence for different $\beta$ value, and noise effect reduces to minimum when $\beta=\pi$.
	
	\textbf{Theorem  1:} Suppose there is a general Grover's algorithm with total $k$ steps, and phase noise is independent in each step. Then the deviation on success probability induced by phase noise will be gradually reduced if Grover's algorithm takes more $\beta$s equal to $\pi$. The noise effect decreases to minimum when Grover's algorithm takes all $\beta$s equal to $\pi$.
	
\begin{figure}[htbp]
		\begin{minipage}{1\linewidth}
			\centering
			\includegraphics[width=0.8\linewidth]{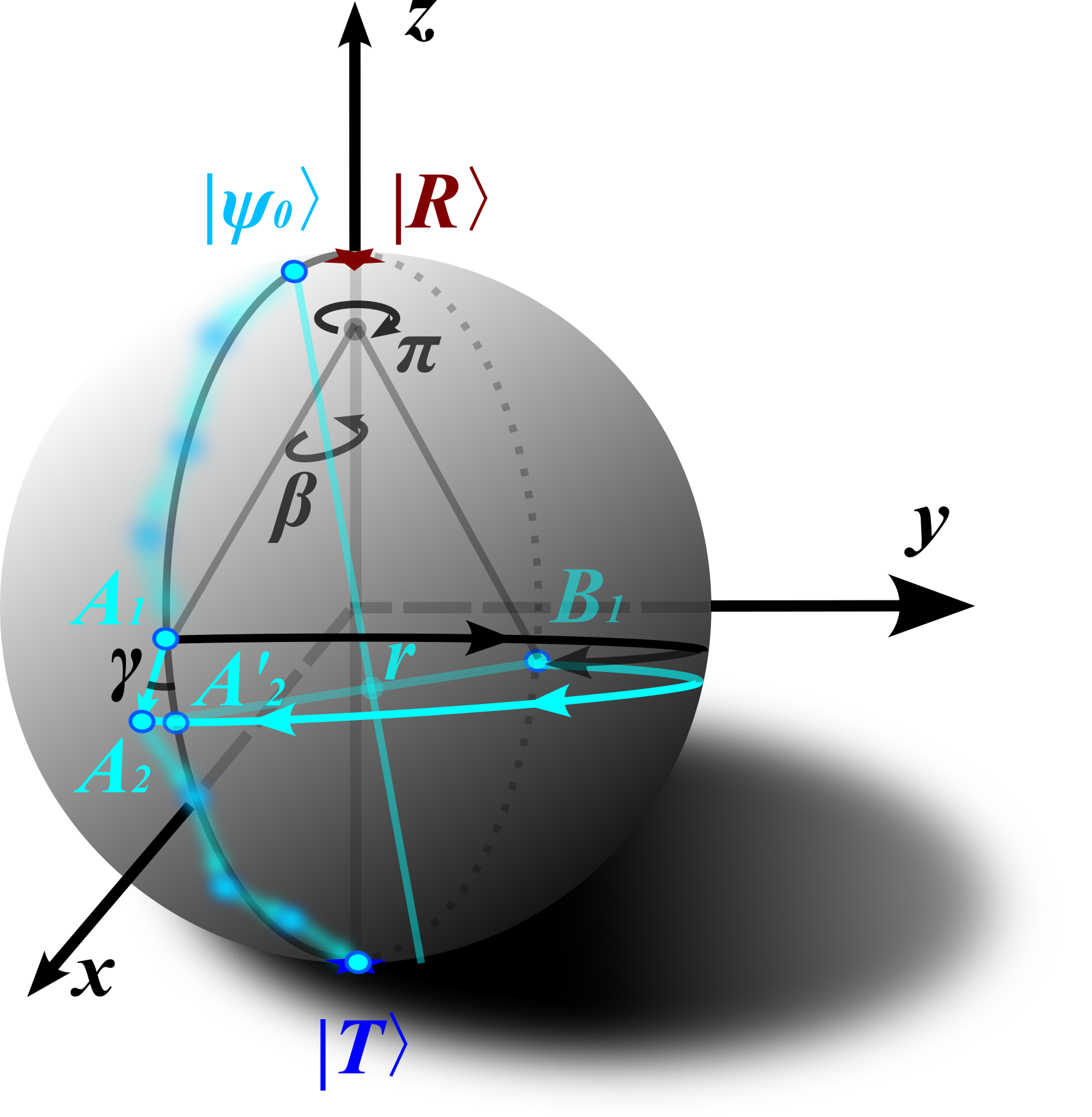}
		\end{minipage}
		
		\caption{\label{one_step}A step of general path which describes $G(-\beta)$ takes $A_1$ to $A_2$. This step can be divided into two stage. First, $S_0$ rotates $A_1$ to $B_1$ about $Z$ axis with $\pi$ angle, so $A_1$ and $B_1$ locates on the same warp. Then $S_r(-\beta)$ rotates $B_1$ to $A_2$ about $|\psi_0\rangle$ with $-\beta$ angle and $r$ radius while this curve intersects the warp at $A_2^\prime$.}
	\end{figure}

	\textbf{Proof:} Consider arbitrary step $\overline{A_1A_2}$ with $-\beta$ of a general path in Fig. \ref{one_step}, where $\overline{A_1A_2}$ is geodesic on Bloch sphere. Rewriting Eq. (\ref{S_matrix}) as
	\begin{eqnarray}
		S_0=&&I-(1-e^{i\pi})|T\rangle\langle T|,\nonumber\\
		S_r(-\beta_1)=&&e^{-i\beta_1}[I-(1-e^{i\beta_1})|\psi_0\rangle\langle\psi_0|],
	\end{eqnarray}
	we see that $S_0$ rotates point $A_1$ to $B_1$ about $z$ axis with $\pi$ angle, then $S_r(-\beta)$ rotates point $B_1$ to $A_2$ about $|\psi_0\rangle$ with $\beta$ angle and $r$ radius while this curve intersects the warp at $A_2^\prime$. Noticing $\overline{A_2A_2^\prime}$ is a small arc we obtain $\overline{A_1A_2}=\overline{A_1A_2^\prime}$ and $\overline{A_1A_2^\prime}\gamma=\overline{A_2A_2^\prime}=r|\beta-\pi|$. So a phase noise $\delta\beta$ takes a first order deviation $\delta\gamma$ on angle $\gamma$, i.e., $\mathcal{O}(\delta\beta)=\mathcal{O}(\delta\gamma)$. Notably, the phase noise $\delta\beta$ discussed in this article is not the incoherent dephasing noise that shrinks the Bloch vector so that the vector always has unity length and lies within the two-dimensional surface of the Bloch sphere. We are interested in deviation on the warped path of $\overline{A_1A_2}$ (this warped path is so approximate to $\overline{A_1A_2^\prime}$ that we do not draw it in the figure):
	\begin{eqnarray}\label{noise}
		\delta~\overline{\mathrm{warped~path~of~}  A_1A_2}=&&\delta(\overline{A_1A_2}\cos\gamma)\nonumber\\
		=&&-\overline{A_1A_2^\prime}\sin\gamma\delta\gamma+\mathcal{O}(\delta\gamma^2)\nonumber\\
		=&&-d\sin\gamma\delta\gamma+\mathcal{O}(\delta\beta^2),
	\end{eqnarray}
	where $\overline{A_1A_2}=\overline{A_1A_2^\prime}$ has been used and $d$ is the length of $\overline{A_1A_2^\prime}$. Clearly, taking $\gamma=0$ can reduce this deviation to minimum. Since noise is independent for each step, we can directly summarize Eq. (\ref{noise}) for every step $i=1,...,k$ to obtain the deviation on whole warped path:
	\begin{equation}\label{whole_noise}
		\delta~\overline{\mathrm{whole~warped~path}}=-\sum_i^k d_i\sin\gamma_i\delta\gamma_i+\mathcal{O}(\delta\beta^2_i).
	\end{equation}
	Indeed, the length of whole warped path is positively correlated with success probability $|\langle T|\psi_f\rangle|^2$.
	So Eq. (\ref{whole_noise}) shows that phase noise has a first order influence on success probability. Notably, taking $\gamma_i=0$ for more steps will gradually reduce the noise effect on success probability, and taking all $\gamma_i=0$ for every step can reduce this effect to minimum. Equally, noise effect will gradually decrease if taking more $\beta_s=\pi$ and it will decrease to minimum if taking all $\beta_s=\pi$ for every step. $\square$
	
	Taking all $\beta_s=\pi$ for every step corresponds to the original Grover algorithm. Although the original Grover's algorithm is most robust, its success probability is unsatisfactory as shown in Fig. \ref{grover_d2p_review}(c). In contrary, the robustness of D2p algorithm is not strongest, but it is deterministic in noiseless environment. The next section will combine the advantages of these two algorithm to obtain the highest success probability under phase noise.
	
	\begin{figure}[htbp]
		\centering
		\begin{minipage}{1\linewidth}
			\centering
			\includegraphics[width=0.8\linewidth]{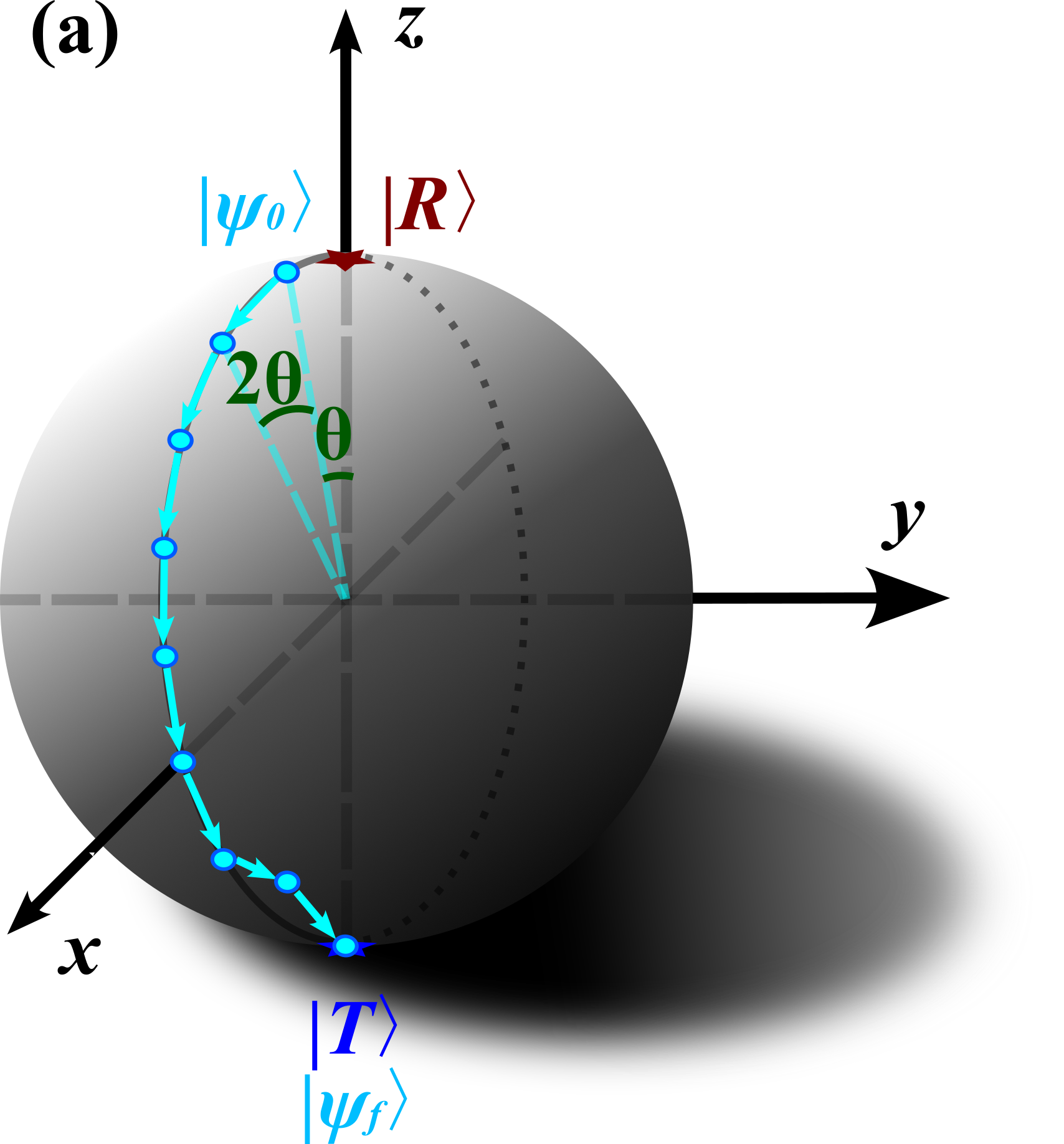}
		\end{minipage}
		
		\begin{minipage}{1\linewidth}
			\centering
			\includegraphics[width=1\linewidth]{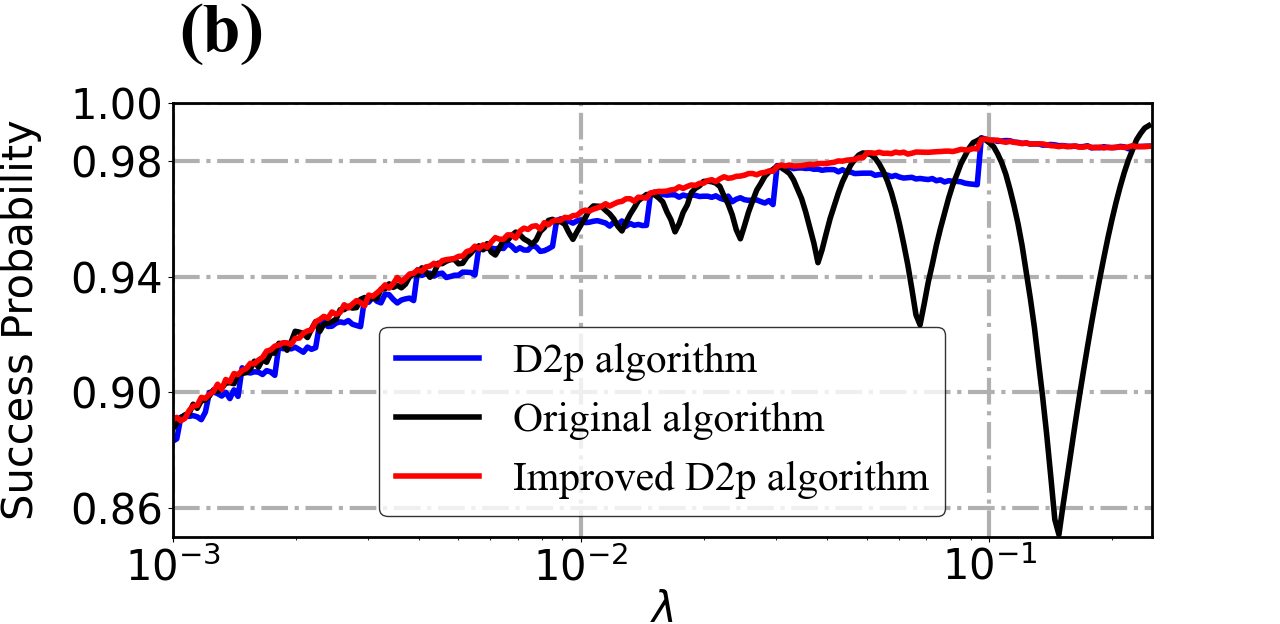}
		\end{minipage}
		
		\begin{minipage}{1\linewidth}
			\centering
			\includegraphics[width=1\linewidth]{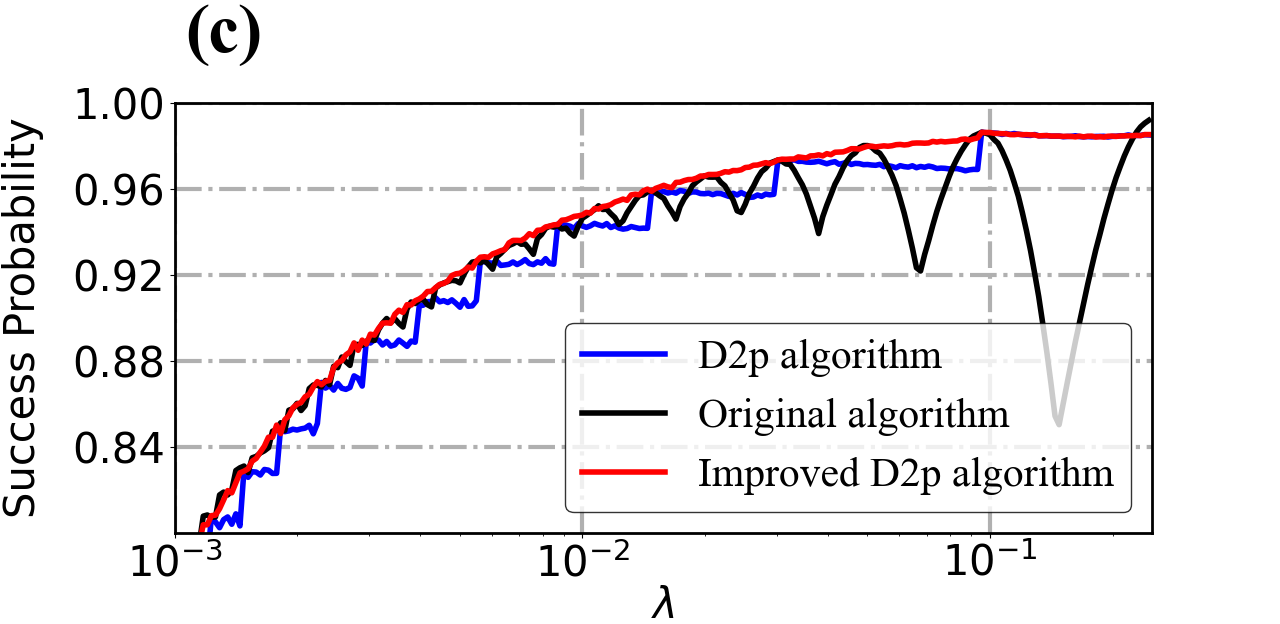}
		\end{minipage}
		
		\caption{\label{improved_d2p}(a) Our improved D2p Grover's algorithm. First, we follow the original Grover's algorithm to rotate initial state $|\psi_0\rangle$ for $k-2=\lceil k_0\rceil-2$ steps. Then we design two particular steps with two phases given by Eq. (\ref{beta_equations}). The final state $|\psi_f\rangle$ will exactly coincide with $|T\rangle$ as D2p in Fig. \ref{grover_d2p_review}(b). (b) We apply the same Gaussian distributed phase noise $\delta\beta=\mathcal{N}(\mu=0,\sigma^2=0.04)$ to original Grover's algorithm, D2p protocol and our improved D2p algorithm. Ten thousand samples are averaged for the data. (c) All works are same to (b) but with the distribution $\mathcal{N}(\mu=0.05,\sigma^2=0.04)$.}
	\end{figure}
	
	\section{Improved D2p Grover's Algorithm}
	
	We improve D2p Grover's algorithm by performing D2p protocol for only two steps while applying the original Grover's search protocol $\beta=\pi$ for all other steps, as shown in Fig. \ref{improved_d2p}(a). Notably, these two D2p steps can be taken at other positions besides the end of algorithm. In \textbf{Appendix A} we show that reserving at different position only makes a negligible effect on success probability and reserving at the end of algorithm can furthest reduce the noise effect.
	
	The step number is $k=\lceil k_0\rceil$. We first apply $G(\pi)$ for $k-2$ steps, then use $G(\beta_1),G(\beta_2)$ for last two steps:
	\begin{equation}\label{psi_final}
		|\psi_f\rangle=G(\beta_2)G(\beta_1)G(\pi)^{k-2}|\psi_0\rangle.
	\end{equation}
	The product $G(\beta_2)G(\beta_1)$ can be written as \cite{roy2022deterministic}
	\begin{equation}
		e^{-\frac{i}{2}(\beta_1+\beta_2)}G(\beta_2)G(\beta_1)=\cos(\phi)I+i\sin(\phi)\sum_{j=x,y,z}\sigma_j n_j,
	\end{equation}
	where $\sigma_{x,y,z}$ are Pauli matrices and
	\begin{eqnarray}
		\cos(\phi)=&&\cos(\frac{\beta_1+\beta_2}{2})+8\lambda(1-\lambda)\sin(\frac{\beta_1}{2})\sin(\frac{\beta_2}{2}),\nonumber\\
		n_x=&&\frac{2\sqrt{\lambda(1-\lambda)}}{\sin(\phi)}\sin(\frac{\beta_1-\beta_2}{2}),\nonumber\\
		n_y=&&\frac{4(1-2\lambda)\sqrt{\lambda(1-\lambda)}}{\sin(\phi)}\sin(\frac{\beta_1}{2})\sin(\frac{\beta_2}{2}),\nonumber\\
		n_z=&&-\frac{(1-2\lambda)}{\sin(\phi)}\sin(\frac{\beta_1+\beta_2}{2}).
	\end{eqnarray}
	The other product $G(\pi)^{k-2}$ can be obtained by diagonalizing $G(\pi)$:
	\begin{equation}\label{G_pi}
		G(\pi)=-X\Lambda X^{-1},
	\end{equation}
	where $X=\frac{1}{\sqrt{2}}(iI+\sigma_x)$ and
	\begin{equation}
		\Lambda=\left(
					\begin{array}{cc}
						e^{i\phi_\lambda} & 0\\
						0 & e^{-i\phi_\lambda}
					\end{array}
				\right)
	\end{equation}
	with $\cos(\phi_\lambda)=1-2\lambda$. The product $G(\pi)^{k-2}$ is simply $(-1)^{k-2}X\Lambda^{k-2}X^{-1}$. Finally the deterministic condition $\langle R|\psi_f\rangle=0$ gives us equations that determine the phase value $\beta_1,\beta_2$:
	\begin{eqnarray}\label{beta_equations}
		&&(\sqrt{1-\lambda}\sin(k-2)\phi_\lambda+\sqrt{\lambda}\cos(k-2)\phi_\lambda)n_x\sin\phi\nonumber\\
		=&&-(\sqrt{1-\lambda}\cos(k-2)\phi_\lambda-\sqrt{\lambda}\sin(k-2)\phi_\lambda)n_z\sin\phi,\nonumber\\
		&&(\sqrt{1-\lambda}\sin(k-2)\phi_\lambda+\sqrt{\lambda}\cos(k-2)\phi_\lambda)n_y\sin\phi\nonumber\\
		=&&-(\sqrt{1-\lambda}\cos(k-2)\phi_\lambda-\sqrt{\lambda}\sin(k-2)\phi_\lambda)\cos\phi.
	\end{eqnarray}
	These equations always have a solution when $\lambda\le\frac{1}{4}$ and the solution is unique as proved in \textbf{Appendix B}.
	
	Now we present the major result of this work as \textbf{Theorem 2} below. With black box (untunable) oracle, we introduce the definition of \textit
	{Deterministic Grover's algorithm}: A type of Grover's algorithm which achieves $100\%$ success probability in noiseless environment.
	
	\textbf{Theorem 2:} Our improved D2p algorithm reaches the upper bound for success probability under phase noise among all possible deterministic Grover's algorithms.
	
	\textbf{Proof:} The success probability of deterministic Grover's algorithm decreases under phase noise. The success probability will reach higher bound if the phase noise is reduced to lower bound. According to \textbf{Theorem 1}, this corresponds to the amount of $\beta=\pi$ steps reaching maximum. Now we prove this maximum amount is $k-2$.
	
	Consider a deterministic Grover's algorithm with $k-1$ steps for $\beta=\pi$, and one step for $\beta\ne\pi$. Applying $G(\pi)$ for $m$ steps, $G(\beta)$ for one step and $G(\pi)$ for $n$ steps, we have
	\begin{equation}
		|\psi_f\rangle=G(\pi)^mG(\beta)G(\pi)^n|\psi_0\rangle.
	\end{equation}
	where $G(\beta)$ and $G(\pi)$ are given by Eqs. (\ref{G_operator}) and (\ref{G_pi}) respectively, and $m+n=k-1$. Rewriting $G(\beta)$ as the summation of real and imaginary parts, we obtain
	\begin{equation}
		|\psi_f\rangle=G(\pi)^m[G_1(\beta,\lambda)+i\sin\beta G_2(\lambda)]G(\pi)^n|\psi_0\rangle.
	\end{equation}
	where $G_1(\beta,\lambda)$ and $G_2(\lambda)$ are real matrices: 
	\begin{align}
		G_1(\beta,\lambda)=&\left(
		\begin{array}{cc}
			(1-\cos\beta)\lambda-1 & (1-\cos\beta)\sqrt{\lambda(1-\lambda)}\\
			(\cos\beta-1)\sqrt{\lambda(1-\lambda)} & (1-\cos\beta)\lambda+\cos\beta
		\end{array}
		\right),\nonumber\\
		G_2(\lambda)=&\left(
		\begin{array}{cc}
			-\lambda &  -\sqrt{\lambda(1-\lambda)} \\
			\sqrt{\lambda(1-\lambda)} & 1-\lambda
		\end{array}
		\right).
	\end{align}
	
	From the imaginary part, deterministic condition $\langle R|\psi_f\rangle=0$ immediately gives solutions $\beta=0,\pi$, since $G(\pi)$ is a real matrix. However, $\beta=0$ gives a trivial operation and $\beta=\pi$ means that we trivially apply the original Grover's algorithm. So $\langle R|\psi_f\rangle=0$ does not have a solution for $\beta$, or there does not exist a deterministic Grover's algorithm with $k-1$ steps for $\beta=\pi$, and one step for $\beta\ne\pi$.
	
	Now we try to construct a deterministic Grover's algorithm with $k-2$ steps for $\beta=\pi$, and two steps for $\beta\ne\pi$. Following the procedure in Eqs. (\ref{psi_final}--\ref{beta_equations}), we obtain the improved D2p Grover's algorithm. $\square$
	
\begin{figure}[htbp]
		\centering
		\begin{minipage}{1\linewidth}
			\centering
			\includegraphics[width=1\linewidth]{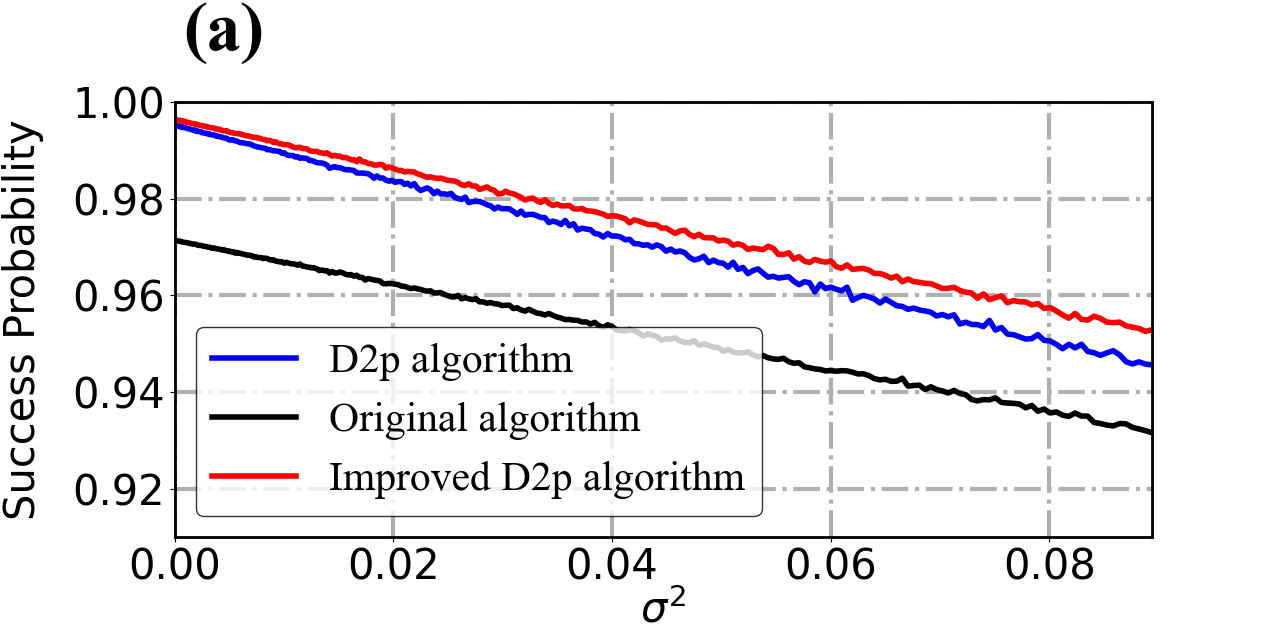}
		\end{minipage}
		
		\begin{minipage}{1\linewidth}
			\centering
			\includegraphics[width=1\linewidth]{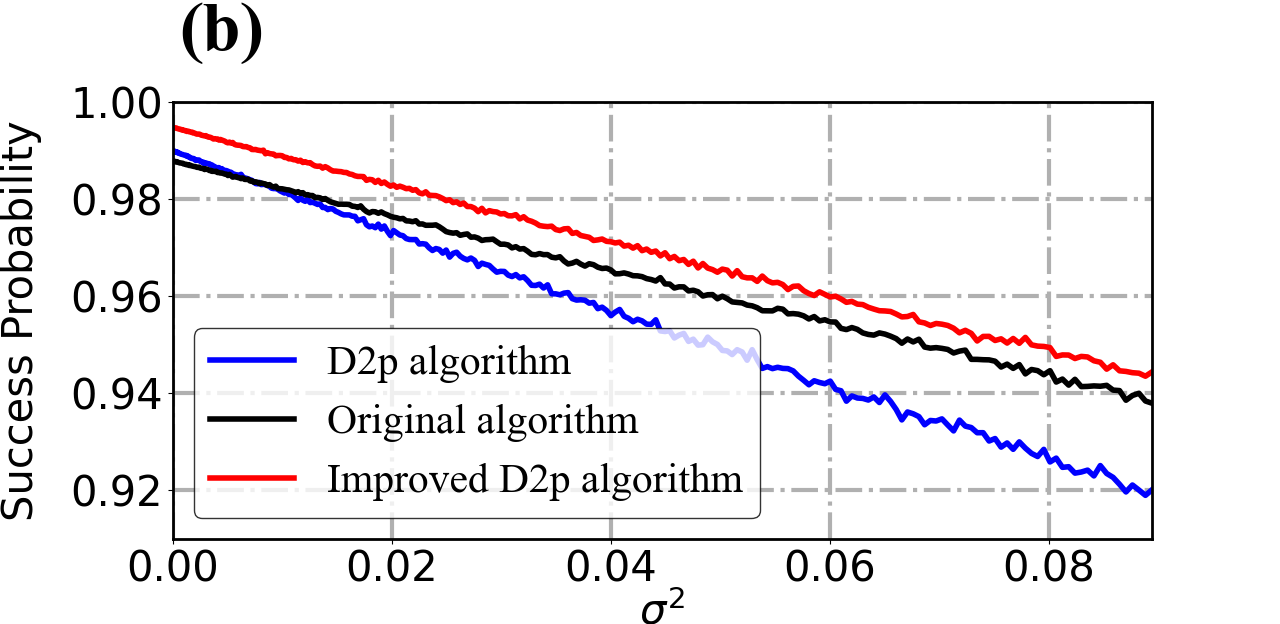}
		\end{minipage}
		
		\caption{\label{variation_effect}(a) We apply the same Gaussian distributed phase noise $\delta\beta=\mathcal{N}(\mu=0.05,\sigma^2)$ to original Grover's algorithm, D2p protocol and our improved D2p algorithm with fixed value $\lambda=0.040$. We collect their success probability under different variation parameter $\sigma^2$. Ten thousand samples are averaged for the data. (b) All are the same to (a) but $\lambda=0.027$.}
	\end{figure}

\begin{figure}[htbp]
		\centering
		\begin{minipage}{1\linewidth}
			\centering
			\includegraphics[width=1\linewidth]{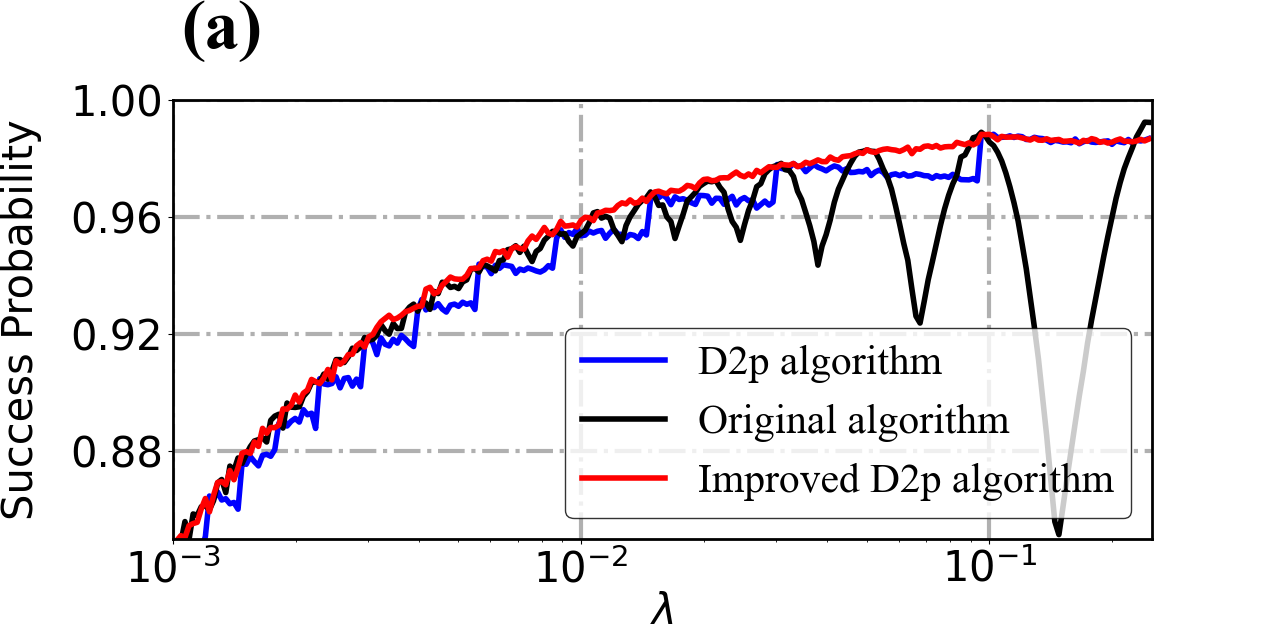}
		\end{minipage}
		
		\begin{minipage}{1\linewidth}
			\centering
			\includegraphics[width=1\linewidth]{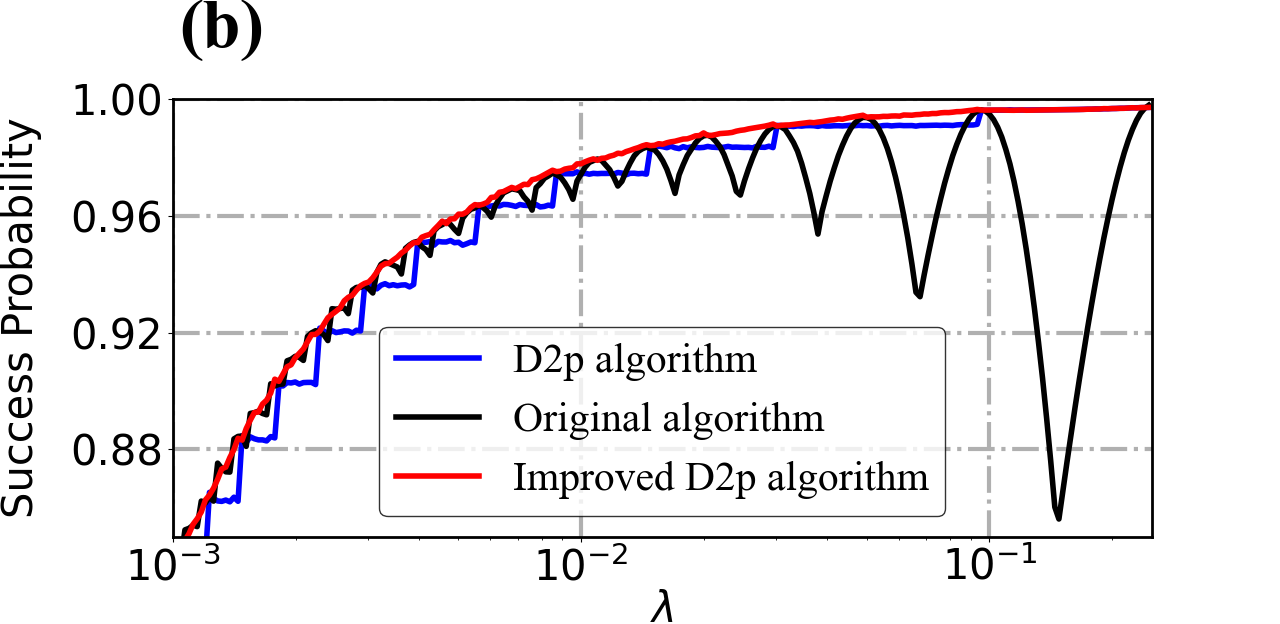}
		\end{minipage}
		
		\caption{\label{different_distribution}(a) We apply the same Poisson distributed phase noise $\delta\beta=\mathcal{P}(0.04)$ to original Grover's algorithm, D2p protocol and our improved D2p algorithm. We collect their success probability with different $\lambda$. Ten thousand samples are averaged for the data. (b) The distribution of noise is changed to uniform $\delta\beta=\mathcal{U}(-0.1,0.2)$. All other conditions are the same to (a).}
	\end{figure}

	Numerical simulations are shown in Fig. \ref{improved_d2p}, \ref{variation_effect}, and \ref{different_distribution}. In Fig. \ref{improved_d2p} (b) and (c), we apply the Gaussian distributed phase noise $\delta\beta=\mathcal{N}(\mu,\sigma^2)$ to the original Grover's algorithm, D2p protocol and our improved D2p algorithm, while the mean value $\mu$ is different for Fig. \ref{improved_d2p} (b) and (c). Fig. \ref{variation_effect} shows that how variation parameter $\sigma^2$ affects the success probability with fixed $\lambda$ when we apply the Gaussian distributed phase noise $\delta\beta=\mathcal{N}(\mu=0.05,\sigma^2)$. In Fig. \ref{different_distribution}, we apply Poisson and uniform distributed noise instead of Gaussian distribution. Simulation results support that our algorithm has the best performance in phase noise environment.

\begin{figure}[htbp]
	\centering
	\begin{minipage}{1\linewidth}
		\centering
		\includegraphics[width=1\linewidth]{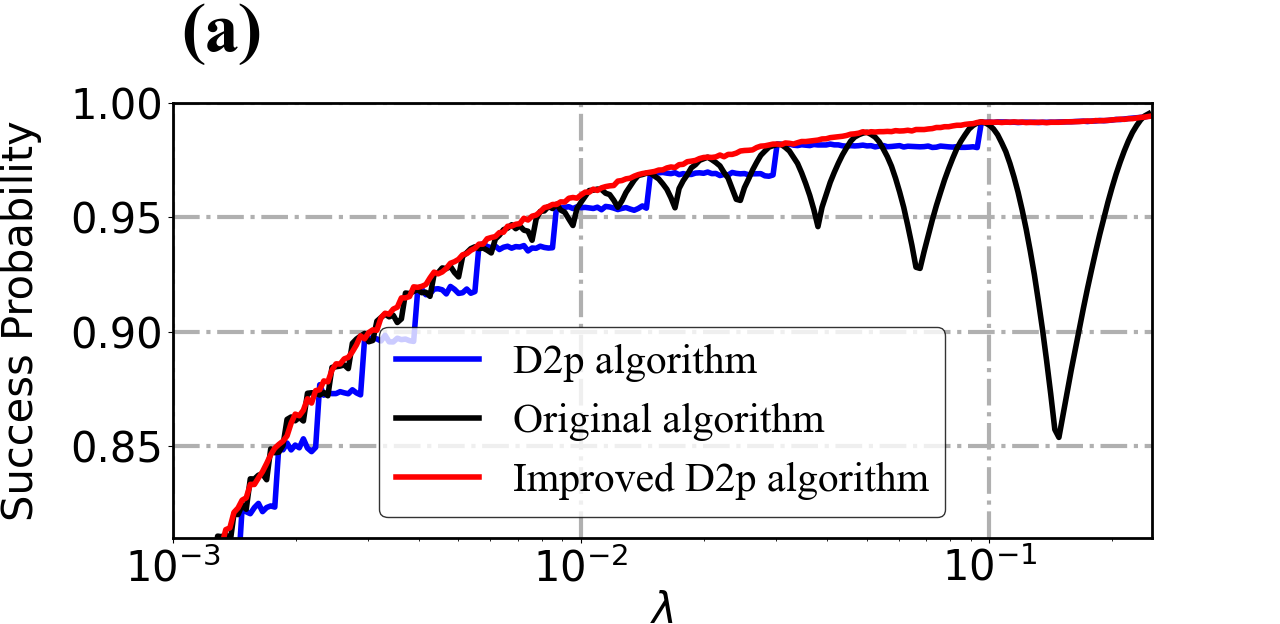}
	\end{minipage}
	
	\begin{minipage}{1\linewidth}
		\centering
		\includegraphics[width=1\linewidth]{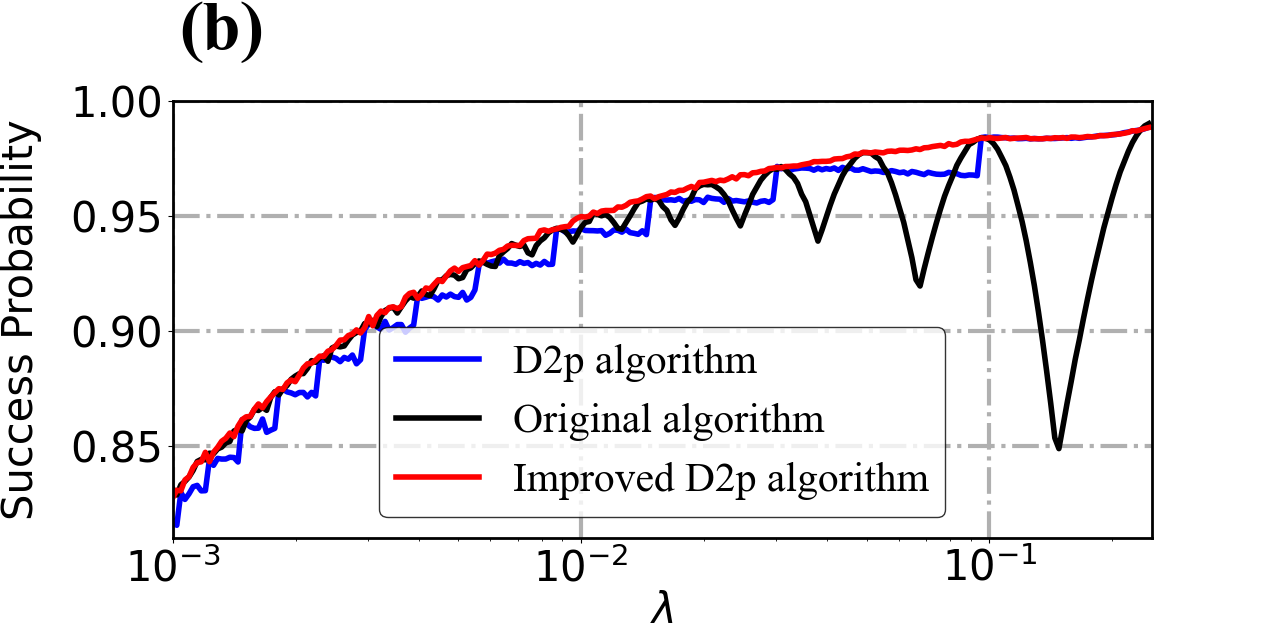}
	\end{minipage}
	
	\caption{\label{oracle_noise}(a) 
		We apply the Gaussian distributed noise $\delta\alpha$ into oracle operator and  $\delta\beta$ into reflection operators where $\delta\alpha$ and $\delta\beta$ obey the same distributed form $\mathcal{N}(\mu=0.03,\sigma^2=0.01)$. The results for original Grover's algorithm, D2p protocol and our improved D2p algorithm are shown in figure. Ten thousand samples are averaged for the data. (b) All works are same to (a) but with the different distribution: $\delta\alpha=\mathcal{N}(\mu=0,\sigma^2=0.04)$ and $\delta\beta=\mathcal{N}(\mu=0.03,\sigma^2=0.01)$.}
	\end{figure}

	We also consider the case that phase noise comes not only into $\beta$, but also into the oracle implementation. Explicitly, oracle operator does not take the form in Eq~.(\ref{S_matrix}) but \cite{roy2022deterministic,long2001grover,toyama2013quantum}
	\begin{equation}
		S_o=\left(
		\begin{array}{cc}
			1 & 0\\
			0 & e^{i(\pi+\delta\alpha)}
		\end{array}
		\right),
	\end{equation}
	where $\delta\alpha$ is the noise. Numerical simulations are shown in Fig. \ref{oracle_noise}. Results show that our algorithm still has the best performance when taking noisy oracle implementation into consideration.
	
	\section{Conclusions} We study geometrical properties of phase noise on Bloch sphere. It gives a new method to study noise effect in Grover's algorithm. Under phase noise, we improve D2p Grover's algorithm to reach the upper bound for success probability. It is a progress for Grover's algorithm working in practical environment.
	
	\begin{acknowledgments}
	We acknowledge the financial support in part by National Natural Science Foundation of China grant No.11974204 and No.12174215.
	\end{acknowledgments}
	
	\appendix
	
	\section{Different positions of two D2p steps}
	
\begin{figure}[htbp]
		\centering
		\begin{minipage}{1\linewidth}
			\centering
			\includegraphics[width=1\linewidth]{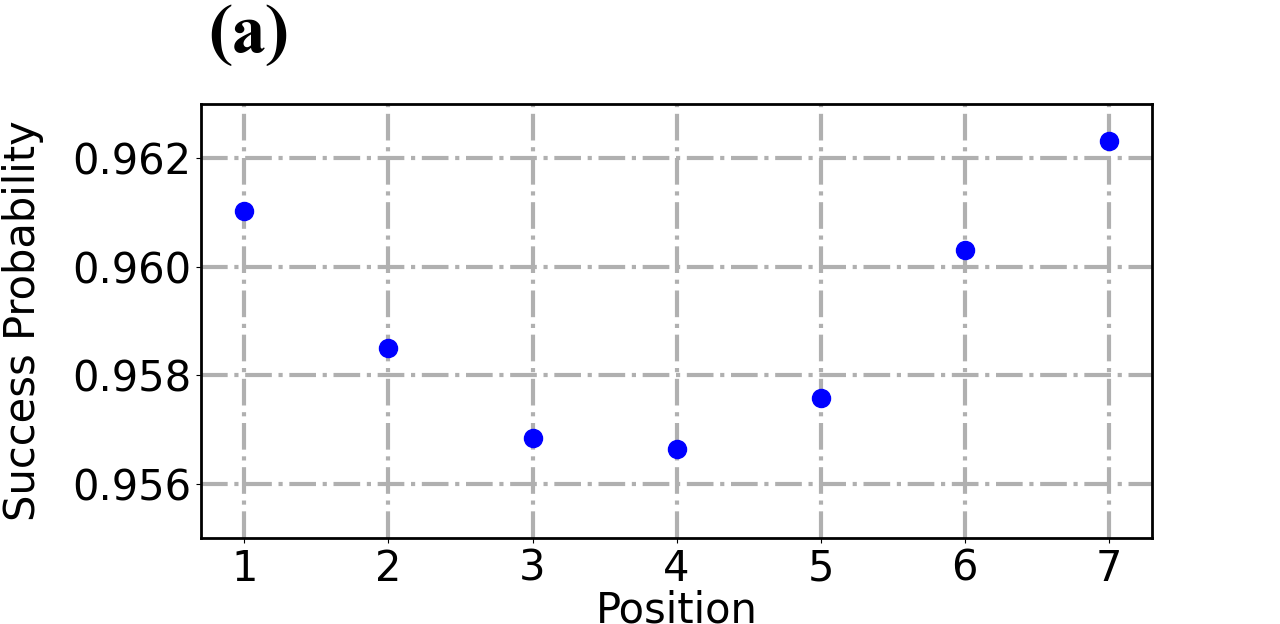}
		\end{minipage}
		
		\begin{minipage}{1\linewidth}
			\centering
			\includegraphics[width=1\linewidth]{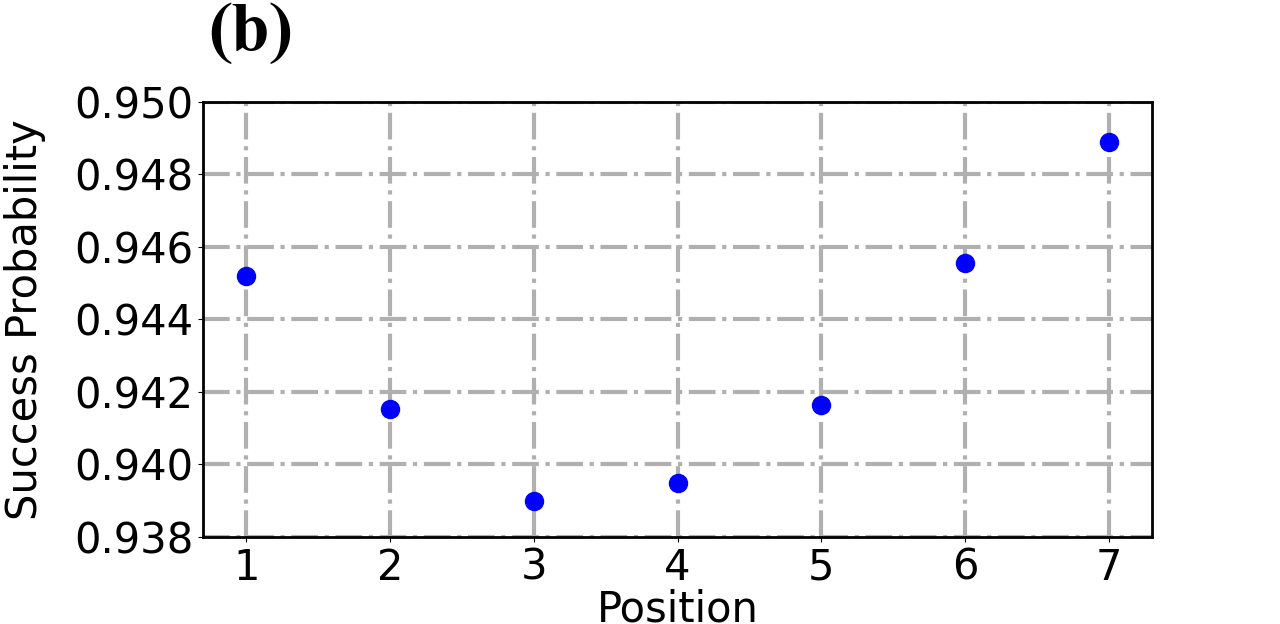}
		\end{minipage}
		
		\caption{\label{different_positions}Taking two D2p steps at different position slightly influences the success probability under phase noise. (a) We set $\lambda=0.01$ and $k=\lceil k_0\rceil=8$ from Eq. (\ref{k_0}) and numerically solve Eq.~(\ref{beta_under_n}) for each position $n$. The position $n=1$ means that we apply two D2p steps at the beginning of algorithm and $n=7$ represents the end of algorithm. We simulate our improved D2p algorithm for every $n$ under phase noise and collect their success probability. The noise obeys Gaussian distribution $\delta\beta=\mathcal{N}(\mu=0,\sigma^2=0.04)$ which is the same as in Fig. \ref{improved_d2p}(b). Fifty thousand samples are average for the data. (b) All works are the same to (a) but the mean value of Gaussian distribution is changed to $\mu=0.05$.}
	\end{figure}

	Here we study the influence that two D2p steps are taken at different positions of D2p algorithm. The step number is $k=\lceil k_0\rceil$. Now we first apply $G(\pi)$ for $n-1$ steps, then use $G(\beta_1),G(\beta_2)$ for two steps, finally apply $G(\pi)$ for $k-n-1$ steps:
	\begin{equation}
		|\psi_f\rangle=G(\pi)^{k-n-1}G(\beta_2)G(\beta_1)G(\pi)^{n-1}|\psi_0\rangle,
	\end{equation}
	where $n=1,2,...,k-1$. Following the same procedure as Eq.~(\ref{psi_final}--\ref{beta_equations}) we obtain the equations that determine the value of $\beta_1$ and $\beta_2$:
	\begin{align}
		&[n_x\sin\phi\cos(k-n-1)\phi_\lambda +n_z\sin\phi\sin(k-n-1)\phi_\lambda]\nonumber\\
		\times&[\sqrt{1-\lambda}\sin(n-1)\phi_\lambda+\sqrt{\lambda}\cos(n-1)\phi_\lambda]\nonumber\\
		=&-[n_z\sin\phi\cos(k-n-1)\phi_\lambda -n_x\sin\phi\sin(k-n-1)\phi_\lambda ]\nonumber\\
		\times&[\sqrt{1-\lambda}\cos(n-1)\phi_\lambda-\sqrt{\lambda}\sin(n-1)\phi_\lambda],\nonumber\\
		&[n_y\sin\phi\cos(k-n-1)\phi_\lambda -\cos\phi\sin(k-n-1)\phi_\lambda]\nonumber\\
		\times&[\sqrt{1-\lambda}\sin(n-1)\phi_\lambda+\sqrt{\lambda}\cos(n-1)\phi_\lambda]\nonumber\\
		=&-[\cos\phi\cos(k-n-1)\phi_\lambda +n_y\sin\phi\sin(k-n-1)\phi_\lambda]\nonumber\\
		\times&[\sqrt{1-\lambda}\cos(n-1)\phi_\lambda-\sqrt{\lambda}\sin(n-1)\phi_\lambda].\label{beta_under_n}
	\end{align}
	These equations are degenerate to Eq.~(\ref{beta_equations}) when taking $n=k-1$. Given $\lambda$ and $n$ we can numerically solve these equations to obtain $\beta_1$ and $\beta_2$, then apply the improved D2p algorithm to get an output under phase noise. Numerical simulation is shown in Fig. \ref{different_positions}. Results show that reserving two D2p steps at different position influences the success probability under phase noise but this influence is negligible compared to Fig. \ref{improved_d2p}. Naturally, we reserve these two steps at the end of our improved D2p algorithm to further reduce the phase noise effect.

	\section{Uniqueness of the solution for Eq.~(\ref{beta_equations})}
\begin{figure}[htbp]
		\centering
		\begin{minipage}{1\linewidth}
			\centering
			\includegraphics[width=0.8\linewidth]{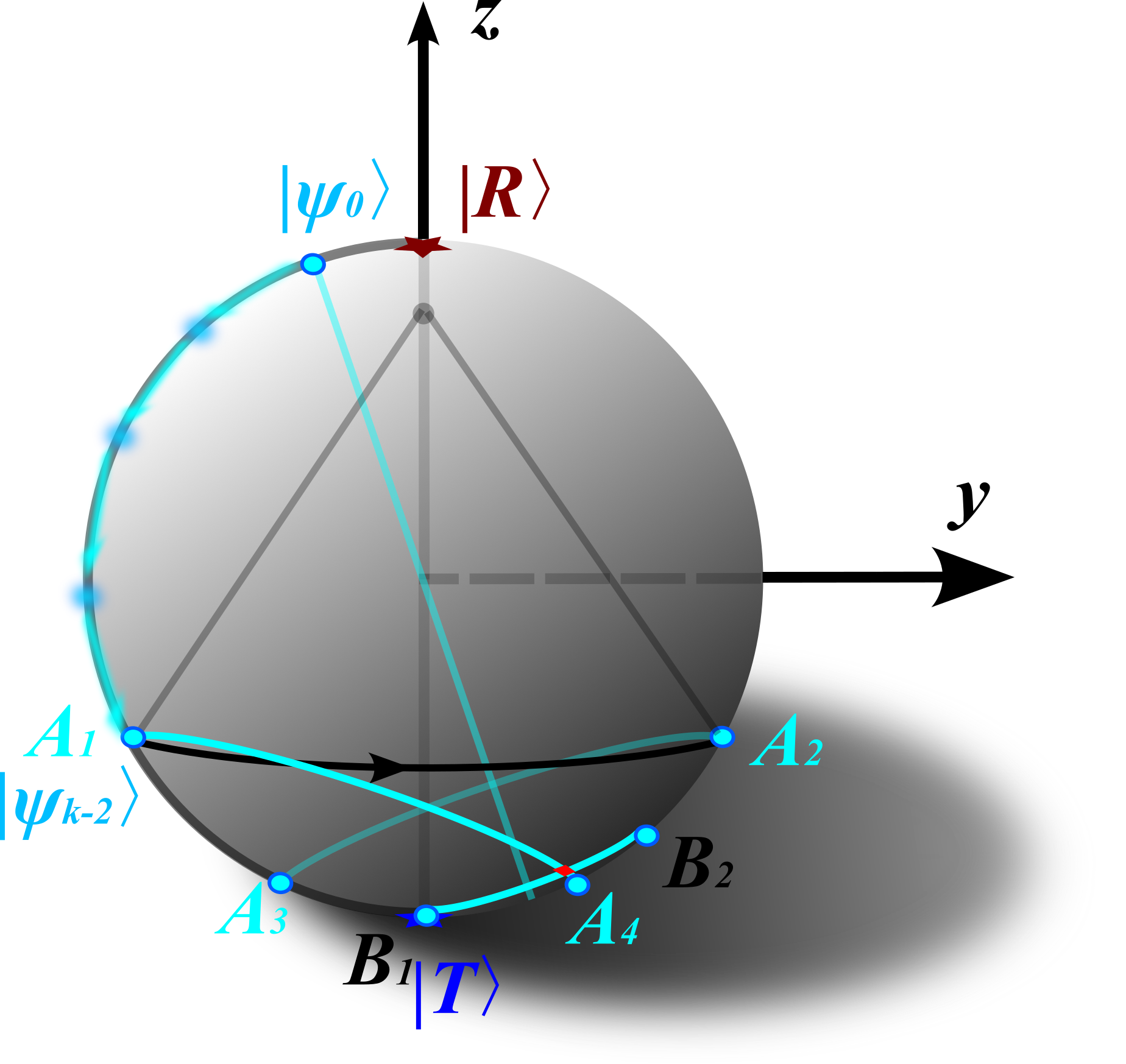}
		\end{minipage}
		
		\caption{\label{uniqueness}Geometrical interpretation for our improved D2p algorithm and its key equation Eq.~(\ref{state_coincide}). For clarity, we do not plot the x-axis. We apply $G(\pi)$ for $k-2$ steps to the initial state $|\psi_0\rangle$ and obtain the state $|\psi_{k-2}\rangle$ which is denoted by the point $A_1$. Without losing any generality, $\beta_1$ is taken to be any value in $[0,~\pi)$ in this figure. According to the right side of Eq.~(\ref{state_coincide}), $S_o$ takes $A_1$ to $A_2$ and $S_r(\beta_1)$ takes $A_2$ to any point on the curve $\overline{A_2A_3}$ where $\overline{A_2A_3}$ locates on the back of sphere. Then $S_o$ takes $\overline{A_2A_3}$ to $\overline{A_1A_4}$ where $\overline{A_1A_4}$ locates on the front of sphere. According to the left side of Eq.~(\ref{state_coincide}), we must take $\beta_2\in(-\pi,~0]$ and $S_r(-\beta_2)$ takes $B_1$ to any point on the $\overline{B_1B_2}$. $\overline{A_1A_4}$ uniquely intersects with $\overline{B_1B_2}$ at the red point.}
	\end{figure}

	It is hard to directly prove the uniqueness by algebraic method since Eq.~(\ref{beta_equations}) is highly nonlinear. Instead, we turn this algebraic problem to a geometrical problem on the surface of Bloch sphere and prove that the solution for this geometrical problem is unique. Now we present our proof with Fig. \ref{uniqueness}.
	
	In our improved D2p algorithm, we first apply $G(\pi)$ for $k-2$ steps to the initial state $|\psi_0\rangle$ and obtain $|\psi_{k-2}\rangle$. Then apply two D2p steps with phase value $\beta_1,~\beta_2$ respectively to obtain the final state $|\psi_f\rangle=|T\rangle$ with an unphysical phase difference:
	\begin{equation}
		|T\rangle=G(\beta_2)G(\beta_1)|\psi_{k-2}\rangle.
	\end{equation}
	According to Eq.~(\ref{S_matrix}) and~(\ref{G_operator}), this is written as
	\begin{align}\label{state_coincide}
		|T\rangle=&S_r(\beta_2)S_oS_r(\beta_1)S_o|\psi_{k-2}\rangle\nonumber\\
		S_r^\dagger(\beta_2)|T\rangle=&S_oS_r(\beta_1)S_o|\psi_{k-2}\rangle\nonumber\\
		S_r(-\beta_2)|T\rangle=&S_oS_r(\beta_1)S_o|\psi_{k-2}\rangle.
	\end{align}
	We should find possible phase values $\beta_1,~\beta_2$ to make the state of both sides coincide for this equation. We first consider the right side of Eq.~(\ref{state_coincide}). Without losing any generality, $\beta_1$ is taken to be any value in $[0,~\pi)$. After applied $S_o$, $S_r(\beta_1)$ and $S_o$, $|\psi_{k-2}\rangle$ is taken to be any position on $\overline{A_1A_4}$. Then we consider the left side of Eq.~(\ref{state_coincide}). $S_r(-\beta_2)$ takes $|T\rangle$ to any point on $\overline{B_1B_2}$ when $\beta_2\in(-\pi,~0]$. Clearly, there is one and only one intersection on $\overline{A_1A_4}$ and $\overline{B_1B_2}$. Corresponding phases $\beta_1,~\beta_2$ on this intersection are the unique solution for Eq.~(\ref{state_coincide}) and~(\ref{beta_equations}). This solution is obtained under $\beta_1\in[0,~\pi)$ and we can also get a solution for $\beta_1\in[-\pi,~0)$ and the difference between these two solutions is trivial: they are symmetric. Now the proof is complete.
	
	\bibliography{refs}
	
\end{document}